\documentclass[twocolumn,floatfix,showpacs,preprintnumbers,amsmath,amssymb]{revtex4}

\usepackage{graphicx}
\usepackage{dcolumn}
\usepackage{bm}

\usepackage[T1]{fontenc}
\usepackage{color}
\usepackage{url}

\input epsf

\begin{document}


\title{Parquet approximation for the $4\times4$ Hubbard cluster}

\author{S.~X.~Yang$^{1}$, H.~Fotso$^{1}$, J.~Liu$^{1}$, T.~A.~Maier$^{2,3}$, K.~Tomko$^{4}$, E.~F.~D'Azevedo$^{2}$, R.~T.~Scalettar$^5$, T.~Pruschke$^{6}$, and M.~Jarrell$^1$} 
\affiliation{$^1$Department of Physics and Astronomy, 
Louisiana State University, Baton Rouge, LA 70803}
\affiliation{$^2$Computer Science and Mathematics Division,Oak Ridge National Laboratory, Oak Ridge, TN 37831}
\affiliation{$^3$Center for Nanophase Materials Sciences, Oak Ridge National Laboratory, Oak Ridge, TN 37831} 
\affiliation{$^4$Ohio Supercomputer Center, Columbus, OH 43212}
\affiliation{$^5$Physics Department, University of California, Davis, CA 95616}
\affiliation{$^6$Department of Physics, University of G\"ottingen,
37077 G\"ottingen, Germany}

\date{\today}

\begin{abstract}
We present a numerical solution of the parquet approximation (PA), a
conserving diagrammatic approach which is self-consistent at both the
single-particle and the two-particle levels. The fully irreducible vertex is
approximated by the bare interaction thus producing the simplest
approximation that one can perform with the set of equations involved in the
formalism. The method is applied to the Hubbard model on a half-filled $4
\times 4$ cluster. Results are compared to those obtained from Determinant
Quantum Monte Carlo (DQMC), FLuctuation EXchange (FLEX), and self-consistent
second-order approximation methods. This comparison shows a satisfactory agreement
with DQMC and a significant improvement over the FLEX or the
self-consistent second-order approximation.
\end{abstract}

\maketitle

\section{Introduction}
Over the past 50 years,
many different techniques have been devised and employed to study strongly correlated
electron systems. Unfortunately, advantages of the successful attempts 
were usually outweight by their limitations. Recently, because of
the progress in computer technology, complex diagramatic approaches have
received increased attention. Although Baym and Kadanoff's $\Phi $
derivability \cite{Baym_Kadanoff,Baym} does not guarantee the physical
validity of a theory, their framework enables the generation of conserving
approximations which are guaranteed to satisfy a variety of Ward identities. For these
reasons, the FLuctuation EXchange (FLEX) approximation \cite{FLEX_Bickers_White,FLEX_Bickers_Scalapino} 
has been intensively studied
over the years. Its major disadvantage however is that it represents a  conserving
approximation at the single-particle level only.
Thus, the physical validity of the approximation appears to be questionable as the vertices are either overestimated or underestimated and the Pauli exclusion principle is not respected properly \cite{Allen Tremblay}. In contrast, the parquet formalism \cite{parquet_old} introduced by 
de Dominicis et al.\ in 1964 is a conserving approximation which is self-consistent also at the two-particle level 
and one may hope that  it resolves at least some of the limitations FLEX has. Unfortunately, it has extremely complicated structure and was, apart from applications to the Anderson impurity model and the 1-D Hubbard model 
with small system size \cite{bickers92,1d_parquet}, hitherto also computationally out of reach.
To circumvent this limitation, Bickers et al.\
introduced the so-called pseudo-parquet approximation \cite{FLEX_Bickers_White} which attempts to improve on the FLEX without introducing the complexity of the full Parquet equations. But this approach fails to properly address the full frequency and momentum dependence of the scattering processes. 
Only very recently, due to the great advance of the parallel computing and the tremendous increase in computer memory,  has it become possible to fully solve this approximation for the first time.

The paper is organised as follows.
In part I we present the
formalism and the resulting equations. In part II, we discuss the algorithm
and the numerical difficulties that arising. In part III, we present
first results obtained from the parquet approximation (PA) for the 2-dimensional
Hubbard model and their comparison to other conserving approximaton methods such as FLEX and self-consistent second-order approximation (SC2nd).
As a benchmark, we compare these results against the Determinant Quantum Monte Carlo (DQMC) which provides a numerically exact result.

\section{Formalism}

\subsection{Vertex functions}

Standard perturbative expansions attempt to describe all the scattering
processes as single- or two-particle Feynman
diagrams. In the single-particle formalism the self-energy describes the
many-body processes that renormalize the motion of a particle in the
interacting background of all the other particles. In the two-particle
context, with the aid of the parquet formalism, one is able to probe the
interactions between particles in greater detail using the so-called vertex
functions, which are matrices describing the two particle scattering
processes. For example, the reducible two-particle vertex $F^{ph}_h(12;34)$
describes the amplitude of a particle-hole pair scattered from its initial
state $\left|3,4\right>$ into the final state $\left|1,2\right>$. Here, $i = 1,2,3,4$ represents a
set of indices which combines the momentum $\mathbf{k}_i$, the matsubara
frequency $i\omega_{n_i}$ and, if needed, the spin $\sigma_i$ and band index 
$m_i$.

In general, depending on how particles or holes are involved in the
scattering processes, one can define three different two-particle scattering
channels. These are the particle-hole (p-h) horizontal channel, the p-h
vertical channel and the particle-particle (p-p) channel. For the Hubbard
model, the spin degree of freedom further divides the particle-particle
channel into triplet and singlet channels while the particle-hole is divided
into density and magnetic channels.

\begin{figure}[!t]
\centerline{
\includegraphics*[width=1.0\hsize]{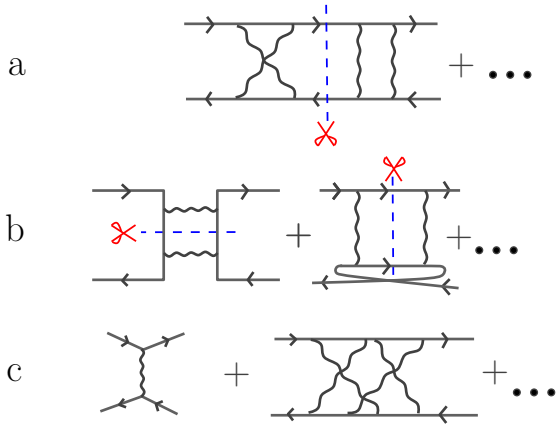}}
\caption{(color online) Different classes of diagrams; the solid line represents 
the single-particle Green's function and the wavy line represents the Coulomb
interaction: here we use the p-h horizontal channel for illustration. (a)
Reducible diagrams: can be separated into two parts by cutting two
horizontal Green's function lines. (b) Irreducible diagrams: can only be
separated into two parts by cutting two Green's function lines in the other
two channels. (c) Fully irreducible diagrams : can not be split in two parts
by breaking two Green's function lines in any channel.}
\label{fig:dia}
\end{figure}

One can further 
discriminate the vertices according to 
their topology.
Starting from the 
reducible vertex $F$ introduced above, we may define the irreducible vertex $\Gamma $
corresponding to the subclass of diagrams in $F$ that can not be separated
into two parts by cutting 
two horizontal Green's function lines. Similarly, the fully
irreducible vertex $\Lambda$ corresponds to the subclass of diagrams in $\Gamma $
that can not be split into two parts by cutting 
two Green's function lines in any channel. An illustration of these different types of vertices is provided in Fig.\ \ref{fig:dia}.

The Pauli exclusion principle produces the so-called crossing symmetries
which in turn yield relationships between these vertices in the
different channels. This enables us to reduce the independent channels
defined for the theory to the particle-particle and the particle-hole
horizontal channels. 


\subsection{Equations}
The parquet formalism assumes the complete knowledge of the fully irreducible vertices 
and provides a set of equations which are self-consistent at both the single- and two-particle
levels. The connection between the single- and two-particle quantities is
through the Schwinger-Dyson equation which connects the reducible vertex $F$
to the self-energy $\Sigma$. It is an exact equation derived from the
equation of motion and has the following form: 
\begin{widetext}
\begin{eqnarray}
\Sigma(P) &=& -\frac{UT^2}{4N} \sum_{P^\prime,Q} \{G(P^\prime) G(P^\prime+Q)
G(P-Q) (F_d(Q)_{P-Q,P^\prime}-F_m(Q)_{P-Q,P^\prime}){\nonumber} \\
&& \;\;\;\;\;\;\;\;\;\;\;\;\;\; + G(-P^\prime) G(P^\prime+Q) G(-P+Q)
(F_s(Q)_{P-Q,P^\prime}+F_t(Q)_{P-Q,P^\prime})\}
\end{eqnarray}
\end{widetext}
where G is the single-particle Green's function, which itself can be calculated from the self-energy using the
Dyson's equation: 
\begin{eqnarray}
G^{-1}(P) = G_0^{-1}(P)\; - \; \Sigma(P)
\end{eqnarray}
Here, the indices $P$, $P^\prime$ and $Q$ combine momentum ${\bf k}$ and Matsubara frequency $i\omega_n$, i.e.\ $P=({\bf k},i\omega_n)$.

The reducible and the irreducible vertices in a given channel are related by
the Bethe-Salpeter equation. It has the following form: 
\begin{eqnarray}
F_{r}(Q)_{P,P^\prime} &=& \Gamma_{r}(Q)_{P,P^\prime} +
\Phi_{r}(Q)_{P,P^\prime} \\
F_{r^\prime}(Q)_{P,P^\prime} &=& \Gamma_{r^\prime}(Q)_{P,P^\prime} +
\Psi_{r^\prime}(Q)_{P,P^\prime}
\end{eqnarray}
where $r=d\;\mbox{or}\;m$ for the density and magnetic channels respectively and $r^\prime=s\;\mbox{or}\;t$ for the singlet and triplet channels, and we are using the vertex ladders which are defined as: 
\begin{eqnarray}
\Phi_{r}(Q)_{P,P^\prime} &\equiv&
\sum_{P^{\prime\prime}}F_{r}(Q)_{P,P^{\prime\prime}}\chi_0^{ph}(Q)_{P^{%
\prime\prime}} \Gamma_{r}(Q)_{P^{\prime\prime},P^\prime} \\
\ \Psi_{r^\prime}(Q)_{P,P^\prime} &\equiv&
\sum_{P^{\prime\prime}}F_{r^\prime}(Q)_{P,P^{\prime\prime}}\chi_0^{pp}(Q)_{P^{%
\prime\prime}} \Gamma_{r^\prime}(Q)_{P^{\prime\prime},P^\prime}
\end{eqnarray}
$\chi_0$ is the direct product of two single-particle Green's functions and
is defined according to the particle-particle or the particle-hole channel.

In a similar manner, the irreducible vertex and the fully irreducible vertex
are related by the parquet equation. This set of equations expresses the fact that the
irreducible vertex in a given channel is still reducible in the other two
channels. The parquet equation has the following form in the different
channels: 
\begin{widetext}
\begin{eqnarray}
\Gamma_d(Q)_{P{P^\prime}} &=& \Lambda_d(Q)_{P{P^\prime}} - {\frac{1 }{2}}%
\Phi_d({P^\prime}-P)_{P,P+Q} - {\frac{3 }{2}}\Phi_m({P^\prime}-P)_{P,P+Q} 
\nonumber \\
&& \;\;\;\;\;\;\;\;\;\;\;\;\;\;\; + \; {\frac{1 }{2}}\Psi_s(P+{P^\prime}%
+Q)_{-P-Q,-P} + {\frac{3 }{2}}\Psi_t(P+{P^\prime}+Q)_{-P-Q,-P} 
\end{eqnarray}
\begin{eqnarray}
\Gamma_m(Q)_{P{P^\prime}} &=& \Lambda_m(Q)_{P{P^\prime}} - {\frac{1 }{2}}%
\Phi_d({P^\prime}-P)_{P,P+Q} + {\frac{1 }{2}}\Phi_m({P^\prime}-P)_{P,P+Q} 
\nonumber \\
&& \;\;\;\;\;\;\;\;\;\;\;\;\;\;\;\; - \; {\frac{1 }{2}}\Psi_s(P+{P^\prime}%
+Q)_{-P-Q,-P} + {\frac{1 }{2}}\Psi_t(P+{P^\prime}+Q)_{-P-Q,-P} 
\end{eqnarray}
\begin{eqnarray}
\Gamma_s(Q)_{P{P^\prime}} &=& \Lambda_s (Q)_{P{P^\prime}} + {\frac{1 }{2}}%
\Phi_d({P^\prime}-P)_{-{P^\prime},P+Q} - {\frac{3 }{2}}\Phi_m({P^\prime}%
-P)_{-{P^\prime},P+Q}  \nonumber \\
&& \;\;\;\;\;\;\;\;\;\;\;\;\;\;\; + \; {\frac{1 }{2}}\Phi_d(P+{P^\prime}%
+Q)_{-{P^\prime},-P} - {\frac{3 }{2}}\Phi_m(P+{P^\prime}+Q)_{-{P^\prime},-P}
\end{eqnarray}
\begin{eqnarray}
\Gamma_t(Q)_{P{P^\prime}} &=& \Lambda_t (Q)_{P{P^\prime}} + {\frac{1 }{2}}%
\Phi_d({P^\prime}-P)_{-{P^\prime},P+Q} + {\frac{1 }{2}}\Phi_m({P^\prime}%
-P)_{-{P^\prime},P+Q}  \nonumber \\
&& \;\;\;\;\;\;\;\;\;\;\;\;\;\;\; - \; {\frac{1 }{2}}\Phi_d(P+{P^\prime}%
+Q)_{-{P^\prime},-P} - {\frac{1 }{2}}\Phi_m(P+{P^\prime}+Q)_{-{P^\prime},-P}
\end{eqnarray}
\end{widetext}
The Bethe-Salpeter and parquet equations are also exact and derived from the categorization of the
Feynman diagrams. 

The above 
discussion of the structure of the parquet formalism is far from being exhaustive and is
merely intended to make the 
paper 
reasonably self-contained. For a more detailed
description of the parquet formalism, we refer the reader to Bickers et al.\ 
\cite{bickers92,bickers98}. Our actual goal is to numerically solve these equations
self-consistently for the Hubbard model on a two dimensional cluster. The
algorithm for this solution is described in the next section.

\section{Algorithm and computational challenge}

\begin{figure}
\centerline{
\includegraphics*[width=0.6\hsize]{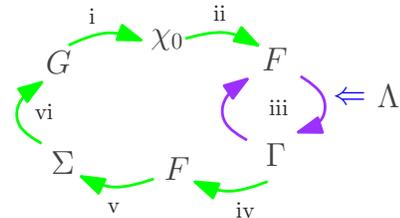}}
\caption{(color online) Schematic illustration for the different steps in solving the parquet approximation equations self-consistently.}
\label{fig:al}
\end{figure}

The set of equations dicussed above are solved self-consistently as
illustrated in the self-consistency loop in Fig. \ref{fig:al}. One starts
with a guess of the single-particle Green's function or self-energy. This can, for example, be
taken from the second-order approximation. The reducible and the irreducible
vertices are also initialized with the bare interaction. The
self-consistency loop can then be described as follows:
\begin{enumerate} 
\item[(i)] first we calculte the bare susceptibility $\chi _{0}$ which is given by the
product of two Green's functions 
\item[(ii)] next this bare susceptibility is used to calculate $F$ through the
Bethe-Salpeter equation 
\item[(iii)] we then proceed with updating the irreducible vertices $\Gamma$ by solving the parquet equation.\newline
This step requires the input of the
fully irreducible vertex $\Lambda $.  In the context of the parquet
approximation which we study here it is taken to be the bare interaction. It however can also be
extracted from some more sophisticated methods. 
\item[(iv)] it is followed by a calculation of the new $F$ through the
Bethe-Salpeter equation 
\item[(v)] this value of $F$ is then used to update the self-energy through the
Schwinger-Dyson equation 
\item[(vi)] the Dyson's equation is solved for the Green's function $G$.%
\newline
\end{enumerate}
This loop is repeated until convergence of the self-energy $\Sigma $ is
achieved within a reasonable criterion.

Unfortunately, this loop becomes unstable when the strength of the Coulomb interaction is increased or the temperature is lowered. 
As we believe that this instability is purely numerical in origin and related to the
iterative nature of the algorithm, we have to 
extend the 
above scheme to account for this problem. For example, one possibility is to start with an
overestimated self-energy and to damp it along with the irreducible vertex
between two iterations according to: 
\begin{eqnarray}  \label{eq:damping1}
\Sigma &=& \alpha_1 \Sigma_{new} + (1-\alpha_1) \Sigma_{old} \\
 \label{eq:damping2}
\Gamma &=& \alpha_2 \Gamma_{new} + (1-\alpha_2) \Gamma_{old}
\end{eqnarray}
where $\alpha_1$ and $\alpha_2$ are some damping parameters.

Another possibility is to rewrite the coupled Bethe-Salpeter and parquet
equations in the form $f({\bf x})=0 \;$ 
 and apply a variant of a Newton's root searching method. Then we can take
advantage of the existing linear solvers such as BiCGS \cite{BICG_Sleijden_Fokkema}, 
GMRES \cite{GMRES_Saad_Schultz} or the Broyden algorithm \cite{Broyden:1988}.

One major advantage that the parquet formalism has over Exact
Diagonalization (ED) or Quantum Monte Carlo (QMC) is that it scales
algebraically with the volume of the system in space-time 
for any choice of parameters including those that lead to a sign problem in QMC. 
The most time-consuming part of the formalism is the solution of
the Bethe-Salpeter and the parquet equations, where the computational time
scales as $O(n_t^4)$ where $n_t = n_c \times n_f$, $n_c$ being the number of
sites on the cluster and $n_f$ the number of Matsubara frequencies.
Although the scaling is better than that of ED or QMC
when the sign problem is severe, one can see that
the complexity quickly grows beyond the capacity of usual desktop computers
with incrasing system size, and large-scale supercomputer systems have to be employed.

Our parallel scheme and our data distribution are based on the realization 
that the Bethe Salpeter equation is the most time-consuming part of our calculation. 
One can easily see that it decouples nicely with respect to the bosonic momentum-frequency index $Q$. 
This enables us to distribute the vertices across processors with respect to this third index 
and to solve the Bethe-Salpeter equation with a local matrix inversion. 
However, this storage scheme puts a limit on the size of the problem that we can address. 
For a node with 2~GBytes of memory, the maximum value of $n_t$ 
that we can use if our variables are complex double precision is about $2500$. 

Unlike the Bethe-Salpeter equation, one can readily observe that the parquet equation does not decouple in terms of the third index. 
Solving this equation requires a  rearrangement of the matrix elements across processors 
and this is the communication bottleneck in the algorithm. 
The rearrangement is necessary to obtain the form of the vertex ladder $\Phi$ or $\Psi$ that is required in the parquet equation. 
For instance, in the $d$ channel, we need $\Phi\left(  P-P^{\prime}\right)  _{P,P+Q}$.
 This form of the vertex ladder is obtained by employing the three-step process described in the following equations:
\begin{eqnarray}  \label{eq:step1}
 \Phi\left(  Q\right)  _{P,P^{\prime}} &\Longrightarrow& \Phi\left(  Q\right)  _{P,P-P^{\prime}} \\
\label{eq:step2}
\Phi\left(  Q\right)  _{P,P-P^{\prime}} &\Longrightarrow& \Phi\left(  P-P^{\prime}\right)  _{P,Q} \\
\label{eq:step3} 
\Phi\left(  P-P^{\prime}\right)  _{P,Q} &\Longrightarrow& \Phi\left(  P-P^{\prime}\right)  _{P,P+Q}
\end{eqnarray}

The first step in this transformation only moves data locally in memory. This does not require much time. 
The second step is actually just a 2D matrix transpose but with matrix elements spreading on many nodes.
 This is where communication across nodes is required. It is achieved by using the standard Message Passing Interface (MPI) collective directives \cite{MPI}. 
The final step is also local and can equally be done very fast. 

\section{Results}

\begin{figure}[t]
\centerline{
\includegraphics*[width=1.0\hsize]{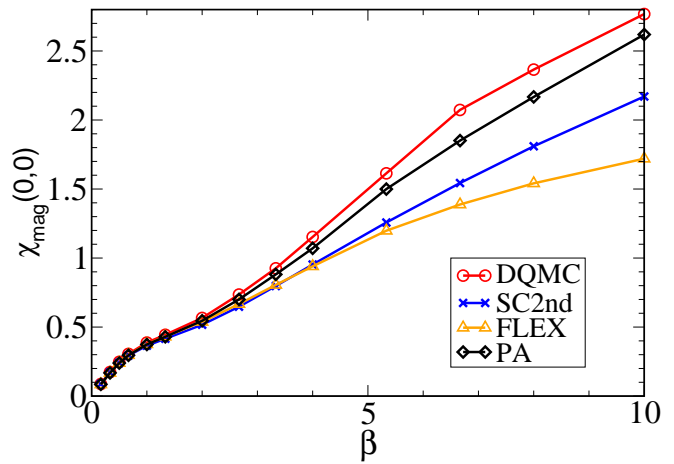}}
\caption{(color online) Uniform susceptibility calculated for different methods as a
function of inverse temperature. While at the high temperature region, all the diagrammatic method results
come close to the DQMC result, the PA shows its advantage clearly in the low temperature region.
}
\label{fig:chi}
\end{figure}

In the following section, we will show the PA results for a $4\times4$
Hubbard cluster at half-filling. The calculations are done for $U=2t$ and
different temperatures. 
The calculations are performed for a finite number of Matsubara frequencies \cite{frequency}.
However, for the observables we calculated, such as the local moment 
and magnetic susceptibility in Fig. \ref{fig:mu} and Fig. \ref{fig:chi},
we performed an extrapolation to an infinite number of frequencies so that the cutoff error in frequency is minimized.
To see how good PA works for the lattice model, we use 
the DQMC result as the benchmark. In the DQMC calculation, $\Delta\tau=1/12$ is used and 
the combined statistical and systematic
errors are smaller than the symbols used. To
further compare PA to other approximations, FLEX and self-consistent second-order results are also included.

\subsection{Single-particle Green function $G\left( \protect\tau\right) $}

\begin{figure}[t]
\centerline{
\includegraphics*[width=1.0\hsize]{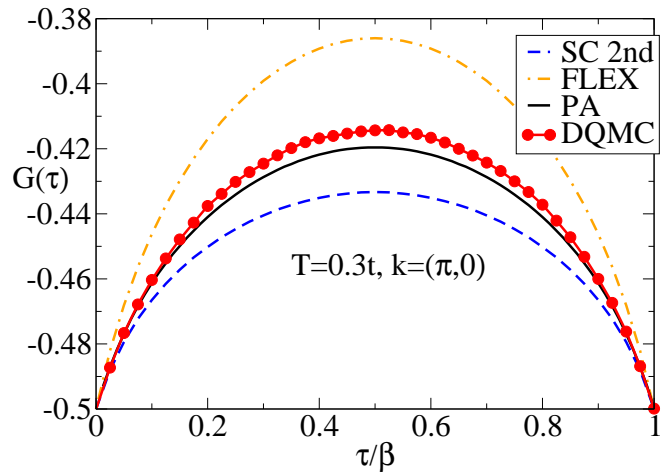}}
\caption{(color online) Single-particle Green function $G\left( \protect\tau \right) $ for
the three diagrammatic approaches and the DQMC. For this temperature, the PA
result (solid line) looks very close to the DQMC one (symbol solid line) as compared to SC second-order (dashed line) or FLEX (dash-dotted line).}
\label{fig:gtau}
\end{figure}

First, one can get a rough idea of how PA improves the accuracy of 
physical observables by comparing the single-particle Green's function from
different levels of approximation. Shown in Fig. \ref{fig:gtau} are $G_{%
\mathbf{k}}\left( \tau \right) $ with $\mathbf{k}=(\pi ,0)$ calculated from
the self-consistent second-order approximation, FLEX, PA and DQMC. 
The parquet result is significantly closer to the DQMC result than the second-order approximation and FLEX results as can readily seen from the figure. 
This confirms the intuition that one would get better results if 
the approximation is made on the vertex which is most irreducible.

\subsection{Unscreened local moment}

\begin{figure}[t]
\centerline{
\includegraphics*[width=0.95\hsize]{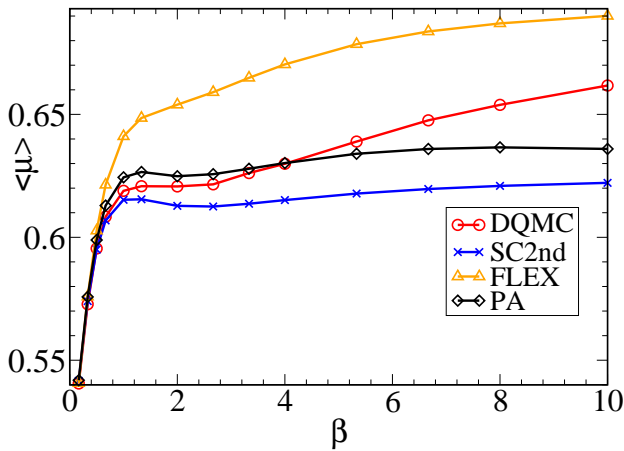}}
\caption{(color online) The inverse temperature dependence of local moment. Among the three
diagrammatic approaches, the PA result comes closest to the DQMC one. }
\label{fig:mu}
\end{figure}

Next we present results for the local magnetic moment defined as 
\begin{eqnarray}
\left< \mu \right> &\equiv &\left< (n_{\uparrow }-n_{\downarrow })^{2} \right> \\
&=&\left< n \right>-2\left< n_{\uparrow }n_{\downarrow } \right>
\end{eqnarray}%
where $\hat{n}_{\sigma }$ denotes the number operator for electrons of spin $%
\sigma $. In the context of a conserving approximation, it can
be re-expressed in terms of the self-energy and the single-particle Green's function as
\begin{eqnarray}
\left< \mu \right>=\left< n \right>-\frac{2T}{U}Tr(\Sigma G)
\end{eqnarray}%
where the trace sums over both the momentum and the frequency degrees of freedom.

The results are shown in Fig. \ref{fig:mu}. Among the three
diagrammatic approaches, the PA result comes closest to the DQMC one. 
If we look more carefully at the DQMC curve, we can find the existence of two humps. 
The hump at $T_{1}\simeq U/2$, which is well reproduced by the PA, 
designates the energy scale for the charge fluctuation, and is directly
related to the suppression of charge double occupancy. The other hump beginning at $%
T_{2}\ll t$ is related to the virtual exchange interaction, $J,$ between
nearby spins. It is believed to be related to the synergism between the
development of the long-range antiferromagnetic correlation and enhancement
of the local moment. As a result, a pseudogap is opened which increases the
entropy of the system \cite{Paiva Scalettar,Moukouri Jarrell}. The
magnitude of $T_{2}$ can be estimated by noticing $J=4t^{2}/U$ for the
strong coupling limit and $t\exp \left( -2\pi t/U\right) $ in the weak
coupling limit \cite{Paiva Scalettar,Hirsch}. Therefore it
basically interpolates between these two limits for that $U=2t$ is
in the intermediate coupling regime. This hump is not well captured by PA. 
The increasing importance of envelop-shape diagram contribution \cite{FLEX_Bickers_White,Allen Tremblay} not included in PA
is responsible for this deviation in the low temperature region.

\subsection{Uniform susceptibility}

Finally, we look at the uniform magnetic susceptibility which is defined as 
\begin{eqnarray}
\chi_{mag}\left( 0,0\right) 
& = & \int_{0}^{\beta}d\tau\left\langle \hat{T}_\tau\,{S}_z\left( \tau\right) {S}%
_z\left( 0\right) \right\rangle \\
& = & \frac{1}{T}\left\langle {S}%
_z^{2}\right\rangle 
\end{eqnarray}
with magnetic moment defined as 
\begin{eqnarray}
\hat{S}_z\left( \tau\right) & = & \frac{1}{N}\sum_{r}\left(
n_{r,\uparrow}\left( \tau\right) -n_{r,\downarrow}\left( \tau\right) \right)
\end{eqnarray}

The $\chi _{mag}$ from different approaches are presented in Fig. \ref%
{fig:chi}. The uniform magnetic susceptibility calculated
from DQMC follows a nearly linear dependence on $\beta .$ This mimics
closely the Curie-Weiss law of weakly interacting moments and implies that
the dominant effect in the system is the short range magnetic fluctuation.
This is consistent with the $\beta $ dependence of the local moment
presented in Fig. \ref{fig:mu}. As the temperature
still dominates over the spin energy scale of the system, it suppresses
the long range fluctuation. 

From this figure, the improvement of PA
over the other two approximations is also easy to see. 
Similar to the local moment, the difference between
results from PA and DQMC at the low temperature region can be explained by the omission of envelop-shape diagrams in PA.

\section{Summary and Outlook}

We have presented the parquet formalism, PA method
and in particular the implementation we use to solve large-sized problem. The preliminary
application of PA on the $4\times4$ Hubbard cluster shows that it can yield
better results than the self-consistent second-order or FLEX calculations. This is the first
step in our work, next we are going to use the parquet formalism in the so-called
Multi-Scale Many-Body (MSMB) approach \cite{msmb}. 
Within MSMB, correlations at different
length scales are treated with different methods. The short length scales
are treated explicitly with QMC methods, intermediate length scales treated
diagrammatically using fully irreducible vertices obtained from QMC and long
length scales treated at the mean field level. 
Note that in this approach the fully irreducible vertex is approximated by a QMC calculation on a small cluster,
while in PA it is approximated by the bare interaction. 
Therefor this approach should provide superior results to the PA.
Another advantage is that it can also avoid
the exponential increase of the computational cost as the system size
increases, and thus can take full advantage of the most up-to-date computer
resources available. We will combine it with the Local Density
Approximation (LDA) to gain some predictive power from the first principle
electronic structure calculation.

\begin{acknowledgments}
We would like to acknowledge the very useful discussion with Gene Bickers and 
John Deisz. SY also acknowlegdes the hospitality and support of the Insitute for Theoretical Physics at the Universit of G\"ottingen, where part of this work has been performed. This work is supported by DOE SciDAC project DE-FC02-06ER25792 
which supports the development of Multi-Scale Many Body formalism and 
codes \cite{msmb} and the DAAD throuhg the PPP exchange program (TP).  SY, HF, KT and MJ are also supported by the NSF PIRE
project OISE-0730290.  This research used resources of the National Center for 
Computational Sciences at Oak Ridge National Laboratory, which is supported by 
the Office of Science of the U.S. Department of Energy under Contract 
No. DE-AC05-00OR22725.

\end{acknowledgments}


\begin{thebibliography}{99}
\bibitem{frequency} We use the periodic boundary condtions in the frequency space for
the ease of implementation. And we have checked that different boundary conditions converge
to the same result as $n_f$ increases.
\end{thebibliography}
\end{document}